\documentclass[aps,prb,twocolumn,showpacs,preprintnumbers,amsmath,amssymb]{revtex4-1}

\usepackage[english]{babel}
\usepackage[utf8]{inputenc}
\usepackage{amssymb}
\usepackage{amsmath}
\usepackage{color}
\usepackage{hyperref}
\usepackage{bbm}
\usepackage{epsfig}

\newcommand{\ket}[1]{|#1\rangle}

\newcommand{\zub}[2]{\langle \! \langle #1 \vert #2 \rangle\!\rangle}

\newcommand{\e}{\varepsilon}
\newcommand{\s}{\sigma}

\newcommand{\dk}{d^\dagger}

\newcommand{\dd}{\partial}

\newcommand{\up}{\uparrow}
\newcommand{\down}{\downarrow}

\newcommand{\beq}{ \begin{equation} } 
\newcommand{\eeq}{ \end{equation} }
\newcommand{\beqa}{\begin{eqnarray}}
\newcommand{\eeqa}{\end{eqnarray}}

\newcommand{\es}{& = &}

\newcommand{\A}{\mathcal{A}}

\newcommand{\fig}[1]{Fig.~\ref{#1}}
\newcommand{\figs}[1]{Figs.~\ref{#1}}
\newcommand{\eq}[1]{Eq.~(\ref{#1})}

\hypersetup{
    colorlinks=true,        
    linkcolor=blue,        
    citecolor=blue,        
    filecolor=blue,      
    urlcolor=blue
}

\begin{document}

\title{Proximity effect on spin-dependent conductance and thermopower\\ of correlated quantum dots}
	   
\author{Krzysztof P. W{\'o}jcik}
\email{kpwojcik@amu.edu.pl}
\author{Ireneusz Weymann}
\email{weymann@amu.edu.pl}
\affiliation{Faculty of Physics, Adam Mickiewicz University, 
			 ul. Umultowska 85, 61-614 Pozna{\'n}, Poland}
\date{\today}

\begin{abstract}
We study the electric and thermoelectric transport properties of correlated quantum dots coupled to 
two ferromagnetic leads and one superconducting electrode.
Transport through such hybrid devices depends on the interplay
of ferromagnetic-contact induced exchange field, superconducting proximity effect
and correlations leading to the Kondo effect.
We consider the limit of large superconducting gap.
The system can be then modeled by an effective Hamiltonian with a particle-non-conserving term
describing the creation and annihilation of Cooper pairs.
By means of the full density-matrix numerical renormalization group method,
we analyze the behavior of electrical and thermal conductances,
as well as the Seebeck coefficient as a function of temperature, dot level position and the strength
of the coupling to the superconductor.
We show that the exchange field may be considerably affected by
the superconducting proximity effect and is generally a function
of Andreev bound state energies. Increasing the coupling to the superconductor
may raise the Kondo temperature and partially restore the exchange-field-split Kondo resonance.
The competition between ferromagnetic and superconducting
proximity effects is reflected in the corresponding temperature and dot level
dependence of both the linear conductance and the (spin) thermopower.
\end{abstract}



\maketitle


\section{Introduction}
\label{sec:intro}


Systems containing quantum dots (QDs) or molecules coupled to different types
of electrodes have been attracting non-decreasing attention for a few decades.
\cite{nato,schon,loss02,andregassen10}
As one of the most interesting phenomena in such systems the Kondo 
effect can be considered,~\cite{kondo} in which the interaction with conduction electrons
gives rise to many-body screening of the localized spin.
\cite{hewson_book}
This results in an additional resonance at the Fermi level
in the local density of states and, consequently, to an enhanced conductance 
through the system for temperatures lower than the Kondo temperature $T_K$. 
\cite{goldhaber-gordon_98,cronenwett_98}

When the electron reservoirs, to which the dot is coupled, exhibit
some correlations, the occurrence of the Kondo
effect is conditioned by the ratio of $T_K$ to the respective
characteristic energy scale of correlated leads.
In particular, in the case of superconducting electrodes,
through multiple Andreev reflections at the quantum dot-superconductor interface,
a Cooper pair carrying two electron charges can be transferred through the dot.
\cite{buitelaarPRL03,jorgensenPRL06,damNature06,hakonenPRL07}
In the Kondo regime this may lead to the conductance enhanced above the unitary limit $2e^2/h$,
provided $T_K$ is larger than the superconducting gap $\Delta$.~\cite{buitelaarPRL02}
If, however, $\Delta > T_K$, the Kondo effect becomes suppressed
and the conductance displays only small side resonances at energies
corresponding to the energy gap.~\cite{hechtJPCM08}
On the other hand, in the case of ferromagnetic leads,
transport properties in the Kondo regime strongly depend on the relative orientation
of the magnetizations of electrodes--the conductance usually drops
when the magnetic configuration switches from the antiparallel into parallel one.
\cite{pasupathy04,barnasJPCM08}
This is related with an exchange field $\Delta\e_{\rm exch}$
that emerges in the parallel configuration and acts in a similar way
as an external magnetic field, splitting the dot level
and thus suppressing the linear conductance.~\cite{pasupathy04,martinekPRL03,hauptmannNatPhys08,gaassPRL11}
It is thus the magnitude of the ferromagnetic-contact-induced exchange field 
that determines the emergence of the Kondo resonance in such systems.
\cite{sindelPRB07,weymannPRB11,wojcikJPCM13}

For quantum dots coupled to both ferromagnetic and 
superconducting leads, transport properties
are conditioned by a sensitive interplay of the exchange field,
correlations leading to the Kondo effect and the superconductivity.
Although transport through such hybrid devices
in the Kondo regime has been recently experimentally measured,~\cite{hofstetterPRL10}
theoretically this problem is still rather unexplored,
although some considerations exist.
\cite{zhuPRB01,fengPRB03,caoPRB04,pengZhang,konig09,konigPRB10,
siqueiraPRB10,wysokinskiJPCM,bocian13,weymann14}
These considerations, however, involved mainly the case
of rather weak tunnel couplings between the dot and external leads,
where the Kondo effect is not fully present and
the effects due to the exchange field are not systematically included.

The goal of the present paper is therefore to provide a 
systematic and reliable analysis of transport properties
of quantum dots with superconducting and ferromagnetic leads in the Kondo regime.
To achieve this goal, we employ the full density-matrix
numerical renormalization group (fDM-NRG) method,
\cite{WilsonRMP75,BullaRMP08,WeichselbaumPRL07,FlexibleDMNRG}
which allows for calculating various linear response transport coefficients in an accurate way.
In particular we focus on the role of Andreev reflection in transport through a quantum
dot coupled to the left and right ferromagnetic leads in a proximity
with the third superconducting lead. We show that the exchange field
due to ferromagnetic leads becomes modified by the coupling to the superconductor
and is determined by the Andreev bound state energies.
This fact is correspondingly reflected in the dependence
of the dot's spectral function and the linear conductance
on the dot level position and temperature.
Moreover, we demonstrate that the effects due to the proximity with the superconductor 
can be also resolved in thermoelectric transport properties.
Thermoelectricity in confined nanostructures, such as quantum dots or molecules,
has recently attracted a lot of attention due to
relatively large values of the figure of merit, \cite{trochaPRB12} which makes such
nanoscale objects interesting for possible future applications.~\cite{hick}
Besides applicatory aspects, it turns out that measuring temperature dependence of the 
thermopower may provide additional information about (Kondo) correlations in the system.~\cite{costiPRB10}
Recently, the Seebeck and spin Seebeck coefficients in Kondo-correlated
quantum dots were studied for nonmagnetic and ferromagnetic leads.
\cite{costiPRB10,rejecPRB12,weymannPRB13,chirlaPRB14}
Here, we will extend these studies to hybrid quantum dots with superconducting
and ferromagnetic electrodes. To determine the thermopower in these hybrid devices,
we assume that there is a temperature gradient between the ferromagnetic leads
and analyze how the proximity effect influences the heat conductance
and (spin) thermopower of the considered system in the Kondo regime.

The paper is organized as follows. In Sec. II we present the theoretical framework 
for our calculations. The model, relevant transport coefficients
and method used in calculations are described therein.
In Sec. III the numerical results on the dot's spectral function 
are presented and the analytical formula for the exchange field is derived.
Section IV is devoted to the discussion of the linear conductance and tunnel magnetoresistance
of the system, while in Sec. V we analyze the thermoelectric transport properties
for different coupling strengths to the superconductor.
Finally, the concluding remarks are given in Sec. VI.
 

\section{Theoretical framework}
\label{sec:model}

\subsection{Model}


\begin{figure}
\centering
\includegraphics[width=0.75\columnwidth]{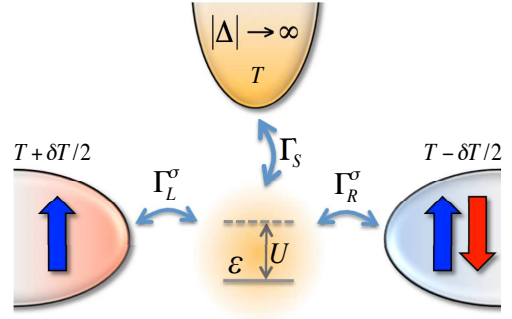}
\caption{\label{fig:schematic}
(color confine) Schematic of the considered system.
Quantum dot, with dot level energy $\e$ and Coulomb correlation $U$,
is connected to two ferromagnetic and one superconducting lead.
$\Gamma_L^\s$ ($\Gamma_R^\s$) describes the spin-dependent coupling between the dot and
the left (right) ferromagnetic lead and $\Gamma_S$ is the coupling to the superconductor.
The magnetizations of ferromagnets can form either 
parallel or antiparallel magnetic configuration, as indicated.
There is a temperature gradient $\delta T$
applied between the two ferromagnetic leads.}
\end{figure}

We consider a single-level quantum dot coupled to two ferromagnetic
leads whose magnetizations can form either parallel (P) or antiparallel (AP) magnetic configuration,
see \fig{fig:schematic}. There is a temperature gradient $\delta T$ applied between
the ferromagnetic leads and the dot is additionally coupled to a superconducting lead.
The Hamiltonian of the system is given by
\beq
H = H_{\rm QD} + H_{\rm F} + H_{\rm S} + H_{\rm TF} + H_{\rm TS}\,,
\label{eq:H}
\eeq
where $H_{\rm QD} =  \sum_\s \e d^\dag_{\s}d_\s + U d^\dag_{\up}d_\up d_{\down}^\dag d_\down$
describes the quantum dot, with $d^\dag_\s$ being the creation operator of 
an electron with spin $\s$ and energy $\e$ in the dot and
$U$ denoting the Coulomb correlations.
The ferromagnetic leads are modeled within the noninteracting quasiparticle
approximation,
$H_{\rm F} = \sum_{\alpha \mathbf{k}\s}  \e_{\alpha \mathbf{k}\s} c^\dag_{\alpha \mathbf{k}\s} c_{\alpha \mathbf{k}\s}$,
where $\alpha = L$ ($\alpha = R$) for the left (right) lead
and $c^\dag_{\alpha \mathbf{k}\s}$ creates an electron of spin $\s$ and momentum $\mathbf{k}$ 
in lead $\alpha$ with the corresponding energy $\e_{\alpha \mathbf{k}\s}$.
The s-wave superconductor is described by,
$H_{\rm S} = \sum_{\mathbf{k}\s}  \xi_{\mathbf{k}\s} a_{\mathbf{k}\s}^\dag a_{\mathbf{k}\s} - \Delta  \sum_{\mathbf{k}}
(a^\dag_{\mathbf{k}\up}a_{-\mathbf{k}\down}^\dag + a_{-\mathbf{k}\down} a_{\mathbf{k}\up})$,
where $\xi_{{\mathbf k} \s}$ denotes the relevant single-particle energy,
$a^\dag_{\mathbf{k}\s}$ is the corresponding creation operator and
$\Delta$ is the superconducting order parameter, which is assumed
to be real and positive.

Finally, the last two terms of the Hamiltonian (\ref{eq:H}) describe tunneling
processes between the dot and ferromagnetic and superconducting leads.
They are respectively given by
\beqa
H_{\rm TF} &=& \sum_{\alpha=L,R}\sum_{\mathbf{k}\s} V_{\alpha \mathbf{k}\s} 
\left(\dk_\s c_{\alpha \mathbf{k}\s} +  c^\dag_{\alpha \mathbf{k}\s} d_\s \right), \\
H_{\rm TS} &=& \sum_{\mathbf{k}\s} V_{S \mathbf{k}\s} 
\left(\dk_\s a_{\mathbf{k}\s} +  a^\dag_{\mathbf{k}\s} d_\s \right),
	\label{HT}
\eeqa
where $V_{\alpha \mathbf{k}\s}$ denotes the tunnel matrix elements between the
dot and ferromagnetic leads, while $V_{S \mathbf{k}\s}$ is the tunnel matrix element
between superconducting electrode and the dot.
The strength of the coupling to ferromagnetic lead $\alpha$
for spin $\s$ is given by $\Gamma_{\alpha}^{\s} = (1+\s p_\alpha)\Gamma$,
where $p_\alpha$ is the spin polarization of ferromagnetic lead $\alpha$,
$p_\alpha = (\Gamma_\alpha^\up - \Gamma_\alpha^\down) / (\Gamma_\alpha^\up + \Gamma_\alpha^\down)$,
and $\Gamma = \Gamma_L + \Gamma_R$, with 
$\Gamma_\alpha = (\Gamma_\alpha^\up+\Gamma_\alpha^\down)/2$
and $\Gamma_\alpha^\s = \pi \rho^{\s}_\alpha V_{\alpha \s}^2$.
Here, $\rho_\alpha^\s$ is the spin-dependent density of states
of lead $\alpha$ and we assumed momentum independent matrix elements
$V_{\alpha \mathbf{k} \s} \equiv V_{\alpha \s}$.
In the following we also assume that the system is symmetric,
$p_L = p_R \equiv p$ and $\Gamma_L = \Gamma_R \equiv \Gamma/2$.
On the other hand, the coupling between the dot and the superconductor
is given by, $\Gamma_S = \pi\rho_S V_S^2$, where $\rho_S$ is the density
of states of the superconductor in the normal state 
and we assumed momentum and spin independent tunnel matrix elements
$V_{S\mathbf{k}\s} \equiv V_S$.

In this paper we focus on the linear-response spin-dependent transport
properties of quantum dots in the proximity with the superconductor.
We assume that the superconducting gap is larger
than the corresponding charging energy of the dot.
It implies that at low temperatures the only processes 
between the dot and superconducting lead are due to the Andreev reflection.
We note that the charging energy in typical quantum dots can range from 
fractions of meV up to a few meV, while the superconducting
energy gap can be as large as a couple of meV.~\cite{nagamatsu01,heinrich13}
Consequently, there are systems where the condition $U < \Delta$ is fulfilled.
Aiming to focus on transport between the two ferromagnets,
we set the electrochemical potential of the superconducting lead to zero
and assume a small symmetric bias between ferromagnetic electrodes.
Then, for symmetric couplings, the net current between the dot and the superconductor vanishes.
In the limit of large $\Delta$, the quantum dot with superconducting lead
can be modeled by the following effective Hamiltonian~\cite{rozhkov,karraschPRB09,mengPRB09}
\beq
H_{\rm QD}^{\rm eff} = H_{\rm QD} + \Gamma_S (\dk_{\up}\dk_{\down} + d_\down d_\up).
\label{eq:Heff}
\eeq
In this Hamiltonian the superconducting degrees of freedom
have been integrated out and the possibility of creating or annihilating Cooper pairs
in the superconductor is now 
included in the last, particle-non-conserving term proportional to $\Gamma_S$.
This effective Hamiltonian can be easily diagonalized and has the following eigenstates:
singly occupied dot states, $\ket{\up}$ and $\ket{\down}$, with energy $\e$
and the two states being combinations of empty ($\ket{0}$) and
doubly occupied ($\ket{\up\down}$) dot states
\beq
\ket{\pm} =  \alpha_\mp \ket{0} \pm \alpha_\pm \ket{\up\down} ,
\eeq
with the coefficients, $\alpha_\pm  = \sqrt{1\pm \delta  / (\delta^2 + \Gamma_S^2)}/\sqrt{2}$,
and $\delta = \e + U/2$ denoting the detuning of the dot level
from the particle-hole symmetry point $\e = -U/2$.
The eigenenergies of the above eigenstates are given by,
$E_\pm = \delta \pm \sqrt{\delta^2 + \Gamma_S^2}$, correspondingly.
The excitation energies of the effective Hamiltonian (\ref{eq:Heff})
result in the following Andreev bound state energies
\beq
  E_{\gamma\eta}^A = \gamma \frac{U}{2} + \eta \sqrt{\delta^2 + \Gamma_S^2}\,,
\eeq
with $\gamma,\eta = \pm$.
They correspond to respective excitations between
the doublet and singlet states of the dot.


\subsection{Transport coefficients}


All the relevant linear-response transport coefficients
can be expressed in terms of the following integral
\beq \label{eq:L}
  L_{n\s} = - \frac{1}{h}\! \int \! d\omega\; \omega^n \, \frac{\dd f(\omega)}{\dd \omega} \, T_\s(\omega),
\eeq
where $f(\omega)$ is the Fermi-Dirac distribution function and
$T_\s(\omega)$ denotes the transmission coefficient.
The spin-resolved linear conductance is then given by
\beq
G_\s = e^2 L_{0\s}.
\eeq

In the case of ferromagnetic leads, depending on the spin relaxation time
in ferromagnets, the voltage drop induced by temperature gradient
can become spin dependent giving rise to spin accumulation,
$\delta V_\s = \delta V +\s V_S$, where $V_S$ is the spin voltage.
One can thus generally distinguish two different situations: (i) the first one
when the spin relaxation is fast enough to assure $V_S=0$ and (ii)
the second one when spin relaxation is slow and $V_S\neq 0$.
Moreover, in the three-terminal setup considered here, 
one needs to be careful about the current which can 
flow to the superconducting lead.~\cite{wysokinskiJPCM}
To prevent the average current from flowing into the superconductor, 
we apply the temperature gradient $\delta T$ symmetrically (see Fig.~\ref{fig:schematic})
and assume that the voltage guaranteeing the absence of the current $J$
induced by temperature gradient is also applied symmetrically,
that is $\mu_l = -\mu_r = \delta V/2$ and $\mu_S = 0$.
Consequently, in the absence of spin accumulation, $V_S=0$,
the thermal conductance is given by
\beq
  \kappa \equiv \left( \frac{\delta J_Q}{\delta T} \right)_{\!\!J=0} = \frac{1}{T} \left[ L_2 - \frac{L_{1}^2 }{ L_{0} }\right],
\eeq
where $L_n = \sum_\s L_{n\s}$.
On the other hand, the Seebeck coefficient is defined as
\beq
S \equiv -\left(\frac{\delta V}{\delta T} \right)_{\!\!J=0} 
	= -\frac{1}{|e|T} \frac{L_{1}}{ L_{0}}.
\eeq

The spin-dependent thermopower in the case of finite spin accumulation,
$V_S\neq 0$ is defined by
\beq
S_\s \equiv -\left(\frac{\delta V_\s}{\delta T} \right)_{\!\!J_\s=0} = -\frac{1}{|e|T} \frac{L_{1\s}}{L_{0\s}},
\eeq
where $J_\s$ denotes the current flowing in the spin channel $\s$.
One can then define the thermopower and the spin thermopower as
\beqa
S_{ac} \es \frac{1}{2} (S_\up + S_{\down}),\\
S_S \es \frac{1}{2} (S_\up - S_{\down}).
\eeqa


\subsection{Method}


To obtain reliable, experimentally testable predictions
for transport properties of correlated quantum dots with ferromagnetic leads
in the proximity with the superconductor, we employ the
full density matrix numerical renormalization group method.
\cite{WilsonRMP75,BullaRMP08,WeichselbaumPRL07,FlexibleDMNRG}
This method allows us to study the dot's local density of states
(dot-level spectral function) as well as
the electric and thermoelectric transport properties
in the full range of parameters in a very accurate way.
In NRG, the conduction band is discretized logarithmically
and the Hamiltonian is mapped onto a tight binding chain with exponentially
decaying hoppings, which can be then diagonalized iteratively.
In our calculations we kept $1024$ states per iteration
and used the Abelian symmetry for the total spin $z$th component.

To perform the analysis, we first applied an orthogonal left-right
transformation to map the effective two-lead Hamiltonian 
to a new Hamiltonian, in which the dot couples only to an even linear combination
of electron operators of the left and right leads,
with a new coupling strength $\Gamma = \Gamma_L+\Gamma_R$.
We note that for left-right symmetric systems, such as considered in this paper,
in the antiparallel configuration the effective coupling is the same for spin-up 
and spin-down electrons. As a result, transport characteristics
are then qualitatively similar to those observed for nonmagnetic systems,
except for a polarization dependent factor.
In the parallel configuration, on the other hand, the effective couplings
do depend on spin polarization of ferromagnets, giving
rise to various interesting effects.

The main quantity we are interested in is the spin-dependent transmission coefficient
\beq
  T_\s(\omega) = \frac{4 \Gamma_L^\s \Gamma_R^\s} {\Gamma_L^\s + \Gamma_R^\s} \pi A_\s(\omega ) ,
\eeq
with $A_\s(\omega)$ being the spin-dependent spectral function of the dot,
$A_\s(\omega) = -\frac{1}{\pi}{\rm Im} G_\s^R(\omega)$, where $G_\s^R(\omega)$
is the Fourier transform of the retarded Green's function of the quantum dot
for spin $\s$. In the parallel and antiparallel magnetic configurations
the spin-resolved transmission coefficient acquires relatively simple form
\beqa
  T_\s^{\rm P}(\omega) &=& (1+\s p) \pi\Gamma A_\s^{\rm P}(\omega), \\
  T_\s^{\rm AP}(\omega) &=& (1-p^2) \pi\Gamma A_\s^{\rm AP}(\omega) ,
\eeqa
respectively. Having determined the transmission, $T_\s (\omega)$,
one can then calculate the integrals $L_{n\s}$, \eq{eq:L},
and find the respective electric and thermoelectric transport coefficients.
However, since in NRG one usually collects the spectral data in logarithmic bins
that are then broadened to obtain a smooth curve, which may introduce 
some errors, we determine the transport coefficients
directly from the discrete, high-quality NRG data.~\cite{rejecPRB12,weymannPRB13}
Nevertheless, when discussing the behavior of the dot spectral function,
to improve its quality and suppress possible broadening artifacts,~\cite{zitko}
in calculations we employ the z-averaging trick
with the number of twist parameters $N_z = 5$.~\cite{oliveira}


\section{Local density of states and exchange field}


\begin{figure}[t]
\centering
\includegraphics[width=0.75\columnwidth]{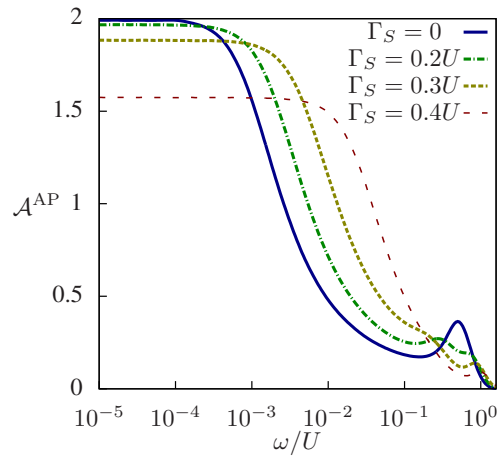}
\caption{\label{fig:AAP}
  (color online) The energy dependence
  of the normalized spectral function in the antiparallel
  magnetic configuration $\mathcal{A}^{\rm AP}$ calculated for $\delta = 0$
  and different couplings to superconducting lead $\Gamma_S$.
  The parameters are: $U=1$, $\Gamma=U/12$, $T=0$ and $p=0.4$.}
\end{figure}


\subsection{Antiparallel configuration}


The normalized spectral function in the antiparallel configuration,
$\A^{\rm AP}(\omega) = \sum_\s \pi \Gamma_\s^{\rm AP} A_\s^{\rm AP}(\omega)$,
is shown in \fig{fig:AAP} for different couplings to the superconductor $\Gamma_S$.
As mentioned above, in the antiparallel configuration the
effective couplings $\Gamma_\s^{\rm AP}$ become spin-independent
and the system behaves as if coupled to nonmagnetic leads.
Consequently, for $\Gamma_S=0$, the spectral function
exhibits the full Kondo resonance at $\omega=0$.
The Kondo temperature for the assumed parameters
and for $\Gamma_S=0$ is, $T_K/U \simeq 1.8\times 10^{-3}$.
There are also two Hubbard resonances, which for $\delta=0$
occur at energies $\omega = \pm U/2$ (note the logarithmic scale in \fig{fig:AAP}).
The proximity of superconducting lead results in gradual suppression of the Kondo effect
with increasing $\Gamma_S$. For finite $\Gamma_S$,
the virtual states for the spin-flip cotunneling processes driving the Kondo effect
are the states $\ket{+}$ and $\ket{-}$, the energy of which greatly depends 
on $\Gamma_S$. This leads to a strong dependence of the Kondo temperature
on the coupling to the superconductor. As can be seen in \fig{fig:AAP},
increasing $\Gamma_S$ generally raises the Kondo temperature.
Moreover, for finite $\Gamma_S$, one can also observe two resonances for larger $\omega$,
which correspond to Andreev bound states of energies $E_{++}^A$ and $E_{+-}^A$,
see \fig{fig:AAP} for e.g. $\Gamma_S/U=0.2$. When the coupling
to the superconductor increases, the energies of the bound states change
and, for larger $\Gamma_S$, the resonance at $\omega=E_{+-}^A$
merges with the Kondo peak, see the curve for $\Gamma_S/U=0.4$ in \fig{fig:AAP}.

The increase of the Kondo temperature for finite $\Gamma_S$ is
due to the fact that the excitation energies from the doublet
state to singlet states $\ket{+}$ and $\ket{-}$ become decreased.
As a consequence, an effective exchange interaction between
the spin in the dot and the conduction electrons
becomes enhanced with increasing $\Gamma_S$.

Another interesting feature visible in \fig{fig:AAP} is the decrease
of the spectral function at the Fermi level with increasing $\Gamma_S$.
In the case of $\Gamma_S=0$, by the Friedel sum rule,~\cite{friedel,hewson_book}
the spectral function $A_\sigma(0)$ at $\omega=0$ 
is given by, $A_\sigma(0) = (\pi\Gamma)^{-1}$,
with $\Gamma_\up=\Gamma_\down\equiv\Gamma$.
To understand the behavior of the spectral function in the presence 
of superconducting lead, let us have a look at the Green's function
$\zub{d_\sigma}{d_\sigma^\dag}_\omega$
of the dot-level, which in the wide band approximation is given by
\begin{equation} \label{Eq:A}
\zub{d_\sigma}{d_\sigma^\dag}_\omega \!=\! \left[ \omega - \e - \Sigma_\sigma + i\Gamma
- \frac{\Gamma_S^2}{ \omega+\e+ \Sigma_{\bar{\sigma}}^S+i\Gamma }\right]^{-1},
\end{equation}
where the self-energies are defined as
\begin{equation*}
\Sigma_\sigma = U \frac{ \zub{d_\sigma n_{\bar{\sigma}}}{d_\sigma^\dag} } {\zub{d_\sigma }{d_\sigma^\dag}}
\;\;\;\;\; {\rm and} \;\;\;\;\;
\Sigma_{\sigma}^S = U \frac{ \zub{d_{\sigma}^\dag n_{\bar{\sigma}}} {d_{\bar{\sigma}}^\dag} }
{\zub{d^\dag_{\sigma} }{d_{\bar{\sigma}}^\dag}}.
\end{equation*}
For the noninteracting case $U=0$, and for $\e=0$, the spectral function
at $\omega=0$ is $A_\sigma(0) = (\pi\tilde{\Gamma})^{-1}$,
with
\beq \label{Eq:Gamma_ren}
 \tilde{\Gamma} = \Gamma\left( 1 + \frac{\Gamma_S^2} {\Gamma^2}\right).
\eeq
Clearly, the height of the spectral function
at the Fermi level decreases with increasing $\Gamma_S$.
The same tendency also holds for the fully interacting case,
see \fig{fig:AAP},
the decrease in $A_\sigma(0)$ is however smaller than in the noninteracting case,
since the denominator in \eq{Eq:Gamma_ren} is larger
due to finite self-energy $\Sigma_{\sigma}^S$, see \eq{Eq:A}.
Note that for $\e=-U/2$ (the particle-hole symmetry point of the Anderson model),
${\rm Re} \{\Sigma_{\sigma}^S(\omega=0)\} \neq U/2$,
contrary to the self-energy $\Sigma_{\sigma}$,
which then fulfills ${\rm Re}\{ \Sigma_{\sigma}(\omega=0)\} = U/2$.


\subsection{Parallel configuration}

\begin{figure}[t]
\centering
\includegraphics[width=1\columnwidth]{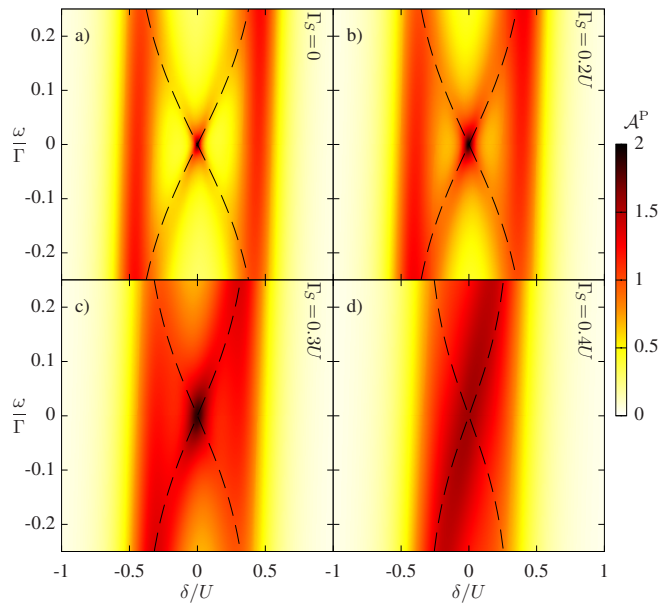}
\caption{\label{fig:AP}
  (color online) The normalized spectral function $\mathcal{A}^{\rm P}$
  in the parallel magnetic configuration as a function
  of energy $\omega$ and level detuning $\delta = \e + U/2$
  for different couplings to superconducting lead:
  (a) $\Gamma_S = 0$, (b) $\Gamma_S/U = 0.2$,
  (c) $\Gamma_S/U = 0.3$, (d) $\Gamma_S/U = 0.4$.
  The dashed lines show the exchange field
  $\Delta\e_{\rm exch}$ obtained from the analytical formula (\ref{eq:Exch}).
  The parameters are the same as in \fig{fig:AAP}.}
\end{figure}

Figure \ref{fig:AP} presents the energy and level detuning dependence
of the normalized spectral function in the parallel magnetic configuration,
$\A^{\rm P}(\omega) = \sum_\s \pi \Gamma_\s^{\rm P} A_\s^{\rm P}(\omega)$.
$\A^{\rm P}(\omega)$ is calculated for a few different values of the coupling
to the superconductor $\Gamma_S$, as indicated in the figure.
By changing the level detuning $\delta$, the occupancy of the dot changes.
For $|\delta| < \sqrt{U^2/4 - \Gamma_S^2}$,
the dot is singly occupied, while for $|\delta| > \sqrt{U^2/4 - \Gamma_S^2}$,
the occupancy is even, i.e. the dot is in state $\ket{+}$
for $\delta < -\sqrt{U^2/4 - \Gamma_S^2}$, and in state $\ket{-}$
for $\delta > \sqrt{U^2/4 - \Gamma_S^2}$.

In the singly occupied regime, for $T<T_K$,
the electronic correlations may give rise to a 
resonance at the Fermi level due to the Kondo effect.
This is indeed what one observes for the antiparallel magnetic configuration, see \fig{fig:AAP}.
However, due to the dependence of tunneling processes on spin,
in the parallel configuration the dot levels for spin-up and spin-down
become renormalized and shift in opposite directions, leading to
a spin splitting of the dot level, $\Delta\e_{\rm exch}$.
This exchange field created by the presence of ferromagnetic leads
suppresses the Kondo effect once $|\Delta\e_{\rm exch}| > T_K$.
Moreover, $\Delta\e_{\rm exch}$ displays a particular dependence on the level detuning $\delta$,
it vanishes for $\delta = 0$  and changes sign at the particle-hole symmetry point.
Although by splitting the dot level the exchange field acts in a similar way as an external magnetic field,
it possesses an extra asset, namely that its magnitude and sign can be
tuned by purely electrical means, i.e. by changing $\delta$ with a gate voltage.

All the above-mentioned features can be clearly visible in \fig{fig:AP}(a),
which presents the spectral function for $\Gamma_S = 0$.
Firstly, the zero-energy spectral function $\mathcal{A}^{\rm P}(0)$
has two maxima broadened by $2\Gamma$ at resonant energies $\delta = \pm U/2$.
Secondly, in the singly occupied dot regime, one observes
the Kondo resonance for $\delta  = 0$, which then becomes split
as the level detuning increases, $|\delta|>0$.
The split Kondo resonance due to the presence of ferromagnetic leads
has already been observed in a number of experiments and is rather well understood.
\cite{pasupathy04,barnasJPCM08,martinekPRL03,hauptmannNatPhys08,
gaassPRL11,sindelPRB07,weymannPRB11,wojcikJPCM13}
Here, we in particular want to analyze how the superconducting proximity
effect affects the exchange field and, thus, the split Kondo resonance.
For finite $\Gamma_S$, the resonant energies are $\delta = \pm\sqrt{U^2/4 - \Gamma_S^2}$.
This implies that the singly occupied regime shrinks with increasing the coupling to the superconductor.
Consequently, the Kondo temperature increases, since the excitation energies from singly occupied
to evenly occupied virtual states, $\ket{+}$ and $\ket{-}$, become lowered.
Moreover, the magnitude of the exchange field also becomes enhanced with increasing $\Gamma_S$.
However, while the increase of $\Delta\e_{\rm exch}$ with $\Gamma_S$ is algebraic,
the $T_K$-dependence on $\Gamma_S$ is rather exponential.
Therefore, for large $\Gamma_S$, the effects due to the proximity
with ferromagnets become eventually overwhelmed by the Kondo correlations.
This is visible in \fig{fig:AP}, where the width of the split Kondo resonances become increased,
till they eventually merge for large $\Gamma_S$, see \fig{fig:AP}(d).
In fact, for $\Gamma_S = 0.4U$, the local moment regime of the dot is relatively narrow
and due to the broadening of the resonant peaks, one observes
only a single low-energy resonance for $\delta = 0$.
The height of this resonance is however lower as compared to the case
of smaller $\Gamma_S$, which indicates that although 
the strong coupling to the superconductor can suppress the 
effects due to the exchange field, it may also destroy the Kondo effect.


\subsection{Perturbative analysis}


\begin{figure}[t]
\centering
\includegraphics[width=0.7\columnwidth]{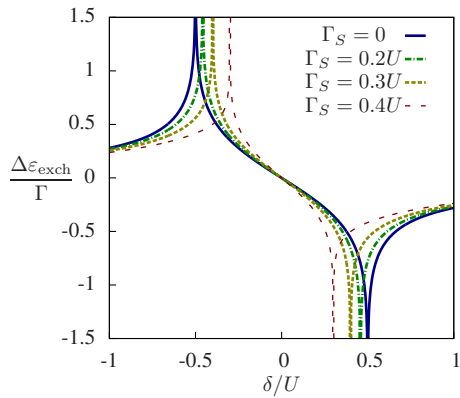}
\caption{(color online) The exchange field in the parallel magnetic configuration
              for a few values of $\Gamma_S$ obtained from \eq{eq:Exch} at $T=0$.
              The other parameters are as in \fig{fig:AAP}.}
\label{fig:exch}
\end{figure}

To estimate the magnitude of the exchange field in the
parallel configuration in the presence of the superconductor,
one can use the second-order perturbation theory to determine
the spin-dependent dot level renormalization $\delta\e_\s$ due to the coupling to ferromagnets.
We thus treat $H_{\rm TF}$ as a perturbation to $H_0 = H_{\rm QD}^{\rm eff} + H_{\rm F}$,
and find that the shift of the level is given by
\beqa
  \delta \e_{\s} 
  &=& - \frac{\Gamma_\s}{\pi} \int \!\! d\omega 
  \left[ \frac{\alpha_+^2 f^-(\omega) } { \omega - E_{-+}^A }  +  \frac{\alpha_-^2 f^-(\omega)} {\omega -  E_{--}^A}   \right]   \nonumber \\
  && - \frac{\Gamma_{\bar{\s}}}{\pi} \int \!\! d\omega 
  \left[ \frac{\alpha_+^2 f(\omega)} { \omega - E_{++}^A }  +  \frac{\alpha_-^2 f(\omega)} {\omega -  E_{+-}^A}   \right] ,
\eeqa
where $f^-(\omega) = 1 - f(\omega)$.
The exchange field can be then obtained from
$\Delta\e_{\rm exch} = \delta\e_\up - \delta\e_\down$ and is given by
\beq \label{eq:Exch}
\Delta \e_{\rm exch} = \frac{2 p \Gamma}{\pi} \frac{\delta}{\sqrt{\delta^2 + \Gamma_S^2}} 
        \left[ \phi(E_{+-}^A) - \phi(E_{++}^A) \right] ,
\eeq
where $\phi(\omega) = {\rm Re} \left\{ \Psi \left( \frac{1}{2} + i \frac{\omega}{2\pi T} \right)\right\}$
and $\Psi(z)$ is the digamma function.
Clearly, the exchange field is a function of the Andreev bound state energies
and can be tuned not only by $\delta$, but also by $\Gamma_S$.
Although $\Delta\e_{\rm exch}$ results directly from the proximity effect with the ferromagnetic leads,
the superconducting proximity effect may considerably affect it.
The formula for the exchange field, \eq{eq:Exch}, can be somewhat simplified at zero temperature
when, $\phi(E_{+-}^A) - \phi(E_{++}^A) = \log | E_{+-}^A / E_{++}^A|$,
while for $\Gamma_S = 0$, one gets,
\cite{martinekPRL03,sindelPRB07,weymannPRB11,wojcikJPCM13}
$\Delta \e_{\rm exch} = \frac{2 p \Gamma}{\pi} \log \left|\frac{\delta-U/2}{\delta+U/2} \right|$.

The exchange field obtained from \eq{eq:Exch}
as a function of level detuning $\delta$ is plotted in \fig{fig:exch}.
The perturbation theory breaks down at resonances, for $|\delta| = \sqrt{U^2/4 - \Gamma_S^2}$,
where the exchange field diverges at $T=0$. We plotted $\Delta\e_{\rm exch}$
in the full range of $\delta$ to present how the resonances move towards the middle
of the Coulomb blockade regime with increasing $\Gamma_S$. Another feature visible
in \fig{fig:exch} is the enhancement of the magnitude of the exchange field
in the singly occupied dot regime with raising the coupling to the superconductor.
$\Delta\e_{\rm exch}$ obtained from formula (\ref{eq:Exch}) is also
shown in \fig{fig:AP} by dashed lines. One can see that the agreement
between the split Kondo resonances visible in the spectral function obtained by NRG
and the estimation for $\Delta\e_{\rm exch}$ based on \eq{eq:Exch} is indeed very good.
For large coupling to the superconductor, however,
the spectral function displays only one broad Kondo resonance and the
splitting is no longer visible due to the broadening of Andreev levels
by the coupling to ferromagnetic leads $\Gamma$.


\section{Linear conductance and tunnel magnetoresistance}


In this section we focus on the role of superconducting proximity effect on
the spin-resolved electric transport coefficients.
In particular, we study the level detuning and temperature dependence
of the linear conductance in the parallel ($G^{\rm P}$)
and antiparallel ($G^{\rm AP}$) magnetic configurations,
as well as the resulting tunnel magnetoresistance,
which is defined as,~\cite{julliere} ${\rm TMR} = G^{\rm P} / G^{\rm AP} - 1$.

\begin{figure}[t]
\centering
\includegraphics[width=1\columnwidth]{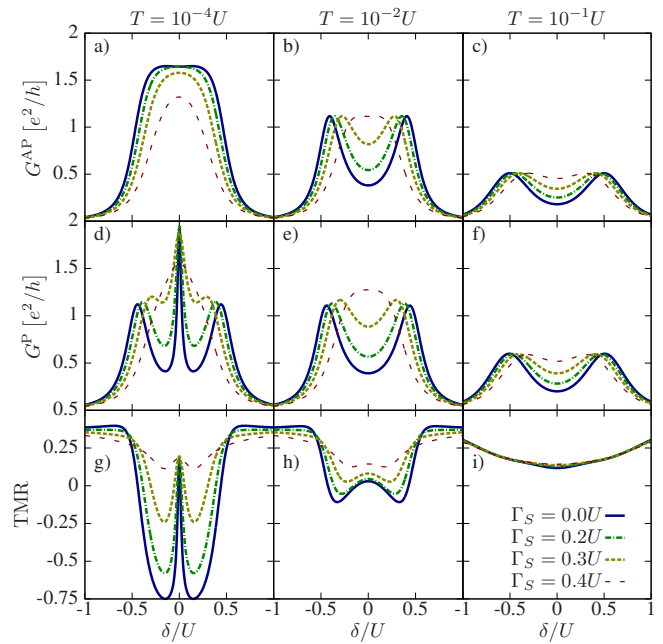}
\caption{(color online) The level detuning dependence of the linear conductance
             in the antiparallel $G^{\rm AP}$ (a)-(c) and parallel $G^{\rm P}$ (d)-(f)
             magnetic configurations as well as the resulting TMR (g)-(i)
             for different couplings to the superconductor and temperatures.
             The left, middle and right columns correspond to $T/U = 10^{-4}$,
             $T/U = 10^{-2}$ and $T/U = 10^{-1}$, respectively.
             The other parameters are as in \fig{fig:AAP}.}
\label{fig:G-E}
\end{figure}


\subsection{Level detuning dependence}


In \fig{fig:G-E} we show the level detuning dependence of the 
linear conductance and the TMR for different temperatures and couplings
to the superconductor. At low temperatures, $T<T_K$, in the antiparallel configuration
there is a Kondo plateau in the singly occupied regime where $G^{\rm AP} = 2(1-p^2)e^2/h$, see \fig{fig:G-E}(a), 
which becomes suppressed with increasing $\Gamma_S$. At intermediate temperatures,
$T/U=10^{-2}$ [\fig{fig:G-E}(b)], for $\Gamma_S = 0$, the Kondo effect is suppressed
since $T>T_K$, however, by increasing the coupling to the superconductor,
one also increases the Kondo temperature and for $\Gamma_S/U=0.4$
there is a single resonance around $\delta = 0$. Note, however,
that this maximum in $G^{\rm AP}$ is mainly due to the fact that the resonant
energies become very close and the two resonant peaks merge due to
the broadening of Andreev levels by the coupling to ferromagnetic leads.
In the case of relatively high temperatures, $T/U = 10^{-1}$, the general dependence
is similar to the previous case, but the conductance is suppressed.

In the parallel configuration, the linear conductance
at low temperatures shows a clear signature of the exchange field
that suppresses the Kondo effect for $\delta\neq 0$, while
for $\delta=0$ the conductance is maximum, $G^{\rm P}=2e^2/h$,
see Fig.~\ref{fig:G-E}(d).
With increasing $\Gamma_S$, the effects due to the exchange field
are effectively decreased and almost completely disappear for $\Gamma_S/U=0.4$,
as explained in Sec. III.
On the other hand, for higher temperatures,
$|\Delta\e_{\rm exch}| < T$, such that thermal fluctuations
smear out the effects due to the exchange field, the dependence of 
$G^{\rm P}$ on $\delta$ is qualitatively similar to that for $G^{\rm AP}$,
with a general tendency that for $T\gtrsim\Gamma$, $G^{\rm P}>G^{\rm AP}$,
compare Figs.~\ref{fig:G-E}(c) and (f).

The difference in $G^{\rm P}$ and $G^{\rm AP}$ gives rise to nonzero TMR
presented in Figs.~\ref{fig:G-E}(g)-(i).
While for $T<T_K$ and $\Gamma_S=0$, the TMR exhibits a highly
nontrivial dependence on level detuning $\delta$,~\cite{weymannPRB11} with increasing
either $\Gamma_S$ or $T$, the TMR dependence on $T$ becomes less dramatic.
First of all, when raising $\Gamma_S$, the effects due to the exchange field
become suppressed and the TMR becomes positive in the whole range
of $\delta$. Moreover, for larger temperatures, the proximity of the superconductor
plays a smaller role and for $T/U = 10^{-1}$, see \fig{fig:G-E}(i),
the TMR is roughly independent of $\Gamma_S$.


\subsection{Temperature dependence}

\begin{figure}[t]
\centering
\includegraphics[width=1\columnwidth]{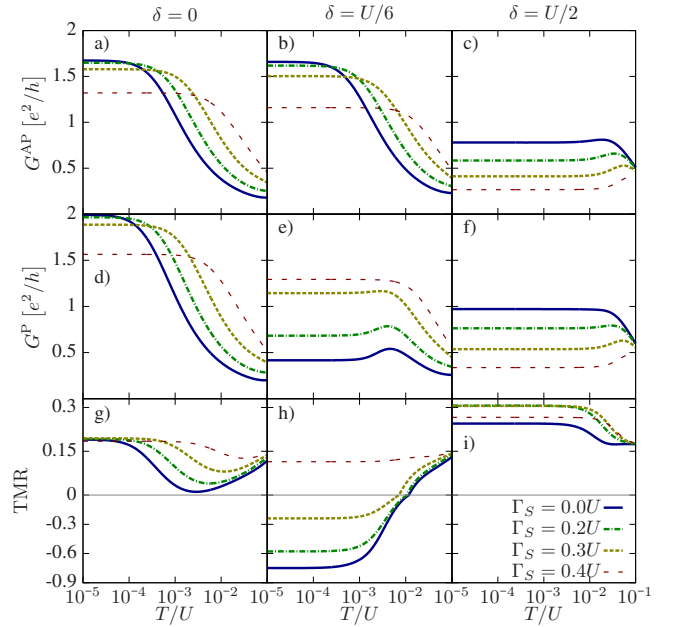}
\caption{(color online) The temperature dependence of $G^{\rm AP}$ (a)-(c), $G^{\rm P}$ (d)-(f)
             and the TMR (g)-(i) for different couplings $\Gamma_S$.
             The left, middle and right columns correspond to $\delta=0$, $\delta=U/6$
             and $\delta = U/2$, respectively. The other parameters are the same as in \fig{fig:AAP}.}
\label{fig:G-T}
\end{figure}

The behavior described above is also visible in the temperature
dependence of the linear conductance and TMR shown in \fig{fig:G-T},
which is calculated for several values of $\delta$ and $\Gamma_S$.
For the particle-hole symmetric case $\delta=0$ presented in the left column
of \fig{fig:G-T}, both $G^{\rm P}$ and $G^{\rm AP}$ exhibit dependence on $T$,
which is typical for quantum dots in the Kondo regime.~\cite{weymannPRB11}
They just differ by a polarization-dependent factor, which in the Kondo regime is equal to $1-p^2$,
and with increasing temperature becomes decreased
for $T\approx T_K$ to raise again once thermally-activated
sequential processes become possible. As a result, the TMR
is given by, ${\rm TMR} = p^2/(1-p^2)$, in the Kondo regime, $T<T_K$, and
in the sequential tunneling regime, $T\gtrsim\Gamma$,
and becomes suppressed for $T\approx T_K$, see \fig{fig:G-T}(g).
With increasing the coupling to the superconductor, these features basically
persist, but the Kondo temperature becomes increased, see Figs.~\ref{fig:G-T}(a) and (d).
In addition, one can observe that the height of the Kondo resonance
is gradually suppressed with increasing $\Gamma_S$.
This can be understood by realizing that the proximity of the superconductor
effectively diminishes the repulsion of electrons in the dot.
Since the Coulomb repulsion is necessary for the Kondo effect to occur,
an increase of $\Gamma_S$ will inevitably lead to the suppression of the
Kondo resonance.

In the Coulomb blockade regime when the particle-hole symmetry
is broken, the exchange field starts playing an important role.
This situation is presented in the middle column of \fig{fig:G-T},
which is calculated for $\delta = U/6$. While the temperature dependence
of $G^{\rm AP}$ is very similar to the case of $\delta=0$, see Figs.~\ref{fig:G-T}(a) and (b),
the linear conductance in the parallel configuration is completely different.
$G^{\rm P}$ is generally suppressed as compared to the
particle-hole symmetric case, which is due to the presence of exchange field.
The Kondo resonance is suppressed and the linear conductance in parallel configuration
displays only a small maximum for temperatures of the order
of the Kondo temperature, see Fig.~\ref{fig:G-T}(e). This results in highly nontrivial dependence
of the TMR on $T$ [Fig.~\ref{fig:G-T}(h)], which now takes large negative values 
for $T<T_K$ and then becomes positive with increasing temperature.
However, when the coupling to superconducting lead increases,
there appears a competition between the exchange field and the superconducting proximity effect,
so that the role of the exchange field becomes diminished,
and the difference in conductances for both magnetic configurations is lowered,
see \figs{fig:G-T}(b) and (e).
Consequently, for relatively strong coupling $\Gamma_S$,
the TMR becomes positive in the whole range of temperatures,
see the case of $\Gamma_S/U=0.4$ in Fig.~\ref{fig:G-T}(h).

The right column of \fig{fig:G-T} presents the case when $\delta=U/2$,
i.e. for $\Gamma_S=0$ the system is on resonance.
With increasing $\Gamma_S$, the resonance moves towards 
the middle of the Coulomb blockade
and the dot becomes occupied by the state $\ket{+}$.
This results in lowering of the linear conductance with increasing $\Gamma_S$,
irrespective of the magnetic configuration of the system, see Figs.~\ref{fig:G-T}(c) and (f).
In fact, when raising the coupling to the superconductor, the transport regime
changes from resonant to cotunneling regime.
As a consequence, the TMR increases with $\Gamma_S$
to reach the value ${\rm TMR} = 2p^2/(1-p^2)$, characteristic of non-spin-flip cotunneling regime.
\cite{weymannPRB05}
However, for higher temperatures, $T\gtrsim\Gamma$, the 
thermally-activated sequential transport dominates and TMR
becomes lowered, reaching ${\rm TMR} = p^2/(1-p^2)$, see Fig.~\ref{fig:G-T}(i).


\section{Seebeck and spin Seebeck coefficients}


\begin{figure}[t]
\centering
\includegraphics[width=0.85\columnwidth]{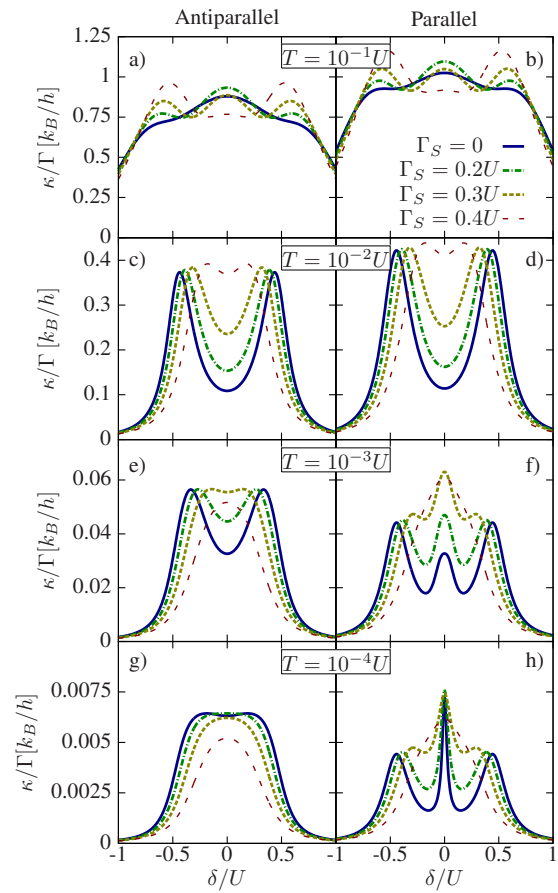}
\caption{\label{fig:k-eps}
(color online) The thermal conductance $\kappa$ as a function
of detuning $\delta$ for the antiparallel (left column)
and parallel (right column) configurations and
for different couplings to the superconductor $\Gamma_S$.
Each row corresponds to different temperature, as indicated.
The parameters are the same as in \fig{fig:AAP}.}
\end{figure}

We now move to the discussion of thermoelectric transport properties of the system.
For this, we assume that there is a temperature gradient $\delta T$
applied to the left and right ferromagnetic leads, see \fig{fig:schematic}.
The formulas for the relevant thermoelectric coefficients are presented in Sec. II.B.
First, we study the influence of the proximity effect
on the thermoelectric coefficients in the case
of no spin accumulation in the leads and then
proceed to the case of finite spin accumulation and 
the analysis of the spin Seebeck effect.

\subsection{Absence of spin accumulation in the leads}

\begin{figure}[t]
\centering
\includegraphics[width=0.85\columnwidth]{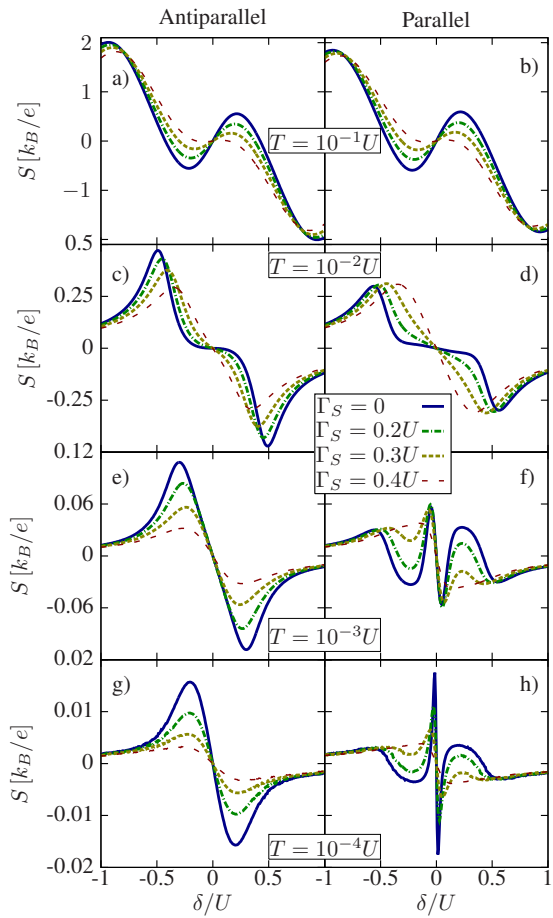}
\caption{\label{fig:S-eps}
(color online) The thermopower as a function
of $\delta$ for the antiparallel (left column)
and parallel (right column) configurations and
for different $\Gamma_S$ and $T$, as indicated.
The other parameters are as in \fig{fig:AAP}.}
\end{figure}

Before analyzing the behavior of the thermopower,
in \fig{fig:k-eps} we show the thermal conductance $\kappa$ as a function
of the level detuning $\delta$ for different temperatures and
couplings to superconducting lead.
The left column corresponds to the antiparallel configuration,
while the right column shows the results in the parallel configuration.
When decreasing temperature, the thermal conductance
becomes generally suppressed, however,
specific shape of its dependence on $\delta$ also changes.
At high temperatures, $T/U=0.1$, $\kappa$
displays a maximum for $\delta=0$, where the particle
and hole processes equally contribute to transport.
This is visible in both magnetic configurations, see \figs{fig:k-eps}(a) and (b),
although $\kappa$ is larger in the parallel configuration 
as compared to the antiparallel one.
With increasing the coupling to the superconductor, 
the maximum changes into local minimum, while
two maxima for $\delta\approx \pm U/2$ develop.
For intermediate temperatures, $T/U=0.01$,
the shape of the $\delta$-dependence of $\kappa$ becomes similar to 
the detuning dependence of the linear conductance, cf. \fig{fig:G-E}.
Similar tendency can be also observed for lower temperatures,
however, while the conductance increases with decreasing $T$,
the thermal conductance becomes suppressed to disappear
completely at zero temperature.
Furthermore, at low temperatures, when $|\Delta\e_{\rm exch}| \gtrsim T$,
the exchange field starts playing an important role and the difference
between both magnetic configurations becomes clearly visible.
When raising $\Gamma_S$, the influence of the exchange field
on transport becomes relatively weakened.
The suppressed thermal conductance in the parallel configuration
in the Coulomb blockade regime becomes then enhanced.
On the other hand, in the antiparallel configuration,
the Kondo effect becomes gradually destroyed and 
$\kappa$ drops in the local moment regime
with increasing $\Gamma_S$, see \figs{fig:k-eps}(g) and (h).

Figure \ref{fig:S-eps} shows the $\delta$-dependence
of the Seebeck coefficient in both magnetic configurations
for different temperatures and couplings $\Gamma_S$.
The behavior of the thermopower is mostly determined 
by the shape of the Kondo peak in the local density of states.
For $\delta=0$, the Kondo resonance is fully symmetric around the Fermi level and,
consequently, the particle and hole currents compensate each other
and the thermopower vanishes. When moving away
from the particle-hole symmetry point, the Seebeck coefficient
becomes nonzero and its sign depends on the relative magnitude
of the particle and hole currents. $S$ is thus an odd function of $\delta$.
At higher temperatures, the behavior of $S$ is qualitatively similar
in both magnetic configurations, see \figs{fig:S-eps}(a) and (b).
The thermopower has a local maximum (minimum) for $0<\delta<U/2$ ($-U/2>\delta>0$).
The differences between $S$ in the parallel and antiparallel configuration
start showing up with lowering temperature, when the exchange field starts playing a role.
In the antiparallel configuration, for $0<\delta<U/2$, the local maximum in $S$ for 
$T/U=0.1$ gradually merges with a local minimum, the position of which 
moves from $\delta \approx U$ towards $\delta=0$ with 
lowering temperature. When increasing the coupling to superconducting lead, the 
thermopower in the antiparallel configuration (and in the parallel configuration
for $T>|\Delta\e_{\rm exch}|$) becomes generally suppressed,
however its qualitative dependence on $\delta$ remains the same.
This is opposite to what we have in the parallel configuration, especially at 
low temperatures, when $T<|\Delta\e_{\rm exch}|$, see \figs{fig:S-eps}(f) and (h).
Then, for $0<\delta<U/2$, $S$ exhibits a local minimum, which becomes sharper
and moves towards $\delta = 0$ with lowering $T$. With increasing $\delta$,
this minimum changes into a local maximum to drop again for
$\delta\approx U/2$. As a consequence, for $\delta>0$,
$S$ changes sign twice in the Coulomb blockade regime.
This is related to the exchange field, which suppresses the
Kondo resonance, once $|\Delta\e_{\rm exch}| > T_K,T$.
Since $\Delta\e_{\rm exch}$ depends strongly on $\delta$,
it leads to the aforementioned behavior of the thermopower
around the particle-hole symmetry point $\delta=0$.
Interestingly, when the coupling to the superconductor is increased,
the $\delta$-dependence of $S$ changes drastically. 
In particular, for large $\Gamma_S$, when the exchange field effects
are suppressed by superconducting proximity effect,
the difference between the two magnetic configurations is decreased,
and $S$ in the parallel configuration behaves similarly to $S$
in the antiparallel configuration. Altogether, this gives rise
to a nontrivial dependence of the low-temperature thermopower
on $\Gamma_S$ in the parallel configuration. The interplay
of the three relevant energy scales: superconducting gap,
exchange field and Kondo temperature
is then clearly revealed, see \fig{fig:S-eps}.

\begin{figure}[t]
\centering
\includegraphics[width=1\columnwidth]{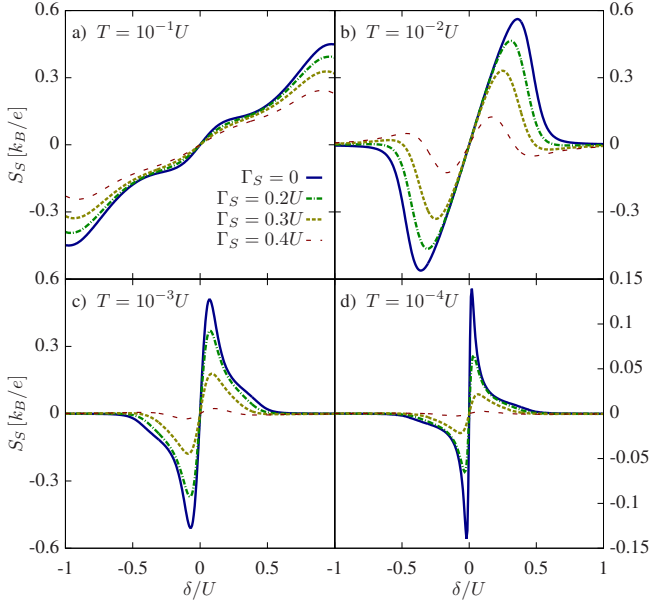}
\caption{\label{fig:SS-eps}
(color online) The spin thermopower
in the parallel configuration as a function
of $\delta$ for different $\Gamma_S$ and for:
(a) $T/U = 10^{-1}$, (b) $T/U = 10^{-2}$,
(c) $T/U = 10^{-3}$ and (d) $T/U = 10^{-4}$.
The other parameters are the same as in \fig{fig:AAP}.}
\end{figure}

\subsection{Finite spin accumulation in the leads}

We now discuss the behavior of the spin thermopower
in the parallel magnetic configuration
in the case of finite spin accumulation in the leads.
Such situation arises when the spin relaxation in the leads is slow
and the shifts of the chemical potential for spin-up and 
spin-down electrons induced by the temperature gradient are not equal.
The $\delta$-dependence of the spin Seebeck coefficient
is presented in \fig{fig:SS-eps} for different temperatures
and couplings to the superconductor.
For $T/U=0.1$, the spin thermopower changes monotonically
with sweeping $\delta$ from $-U$ to $U$, see \fig{fig:SS-eps}(a).
Finite coupling to the superconductor leads only to the suppression of $S_S$.
For smaller temperatures, however, $S_S$ exhibits a maximum
(minimum) for $\delta>0$ ($\delta<0$), see the case of $T/U=0.01$ in \fig{fig:SS-eps}.
This maximum is still present when the coupling to the superconductor becomes stronger,
while its position moves towards the middle of the Coulomb blockade with increasing $\Gamma_S$.
Moreover, for relatively strong coupling to the superconductor, $\Gamma_S/U=0.4$, one can see
that the $\delta$-dependence of the spin thermopower has changed qualitatively.
Now, $S_S$ for $\delta>0$ exhibits a sign change in the Coulomb blockade regime,
which was not present in the case of $\Gamma_S=0$.
When further decreasing the temperature, the maximum in $S_S$
for positive detuning moves towards the particle-hole symmetry point $\delta=0$
and its magnitude becomes suppressed, see \figs{fig:SS-eps}(c) and (d).
For given temperature, increasing the strength of the coupling to superconducting lead,
results in a large suppression of the spin thermopower.
The spin Seebeck coefficient for $T\lesssim T_K$ becomes 
almost fully suppressed in the case of strong coupling to the superconductor.

\begin{figure}[t]
\centering
\includegraphics[width=1\columnwidth]{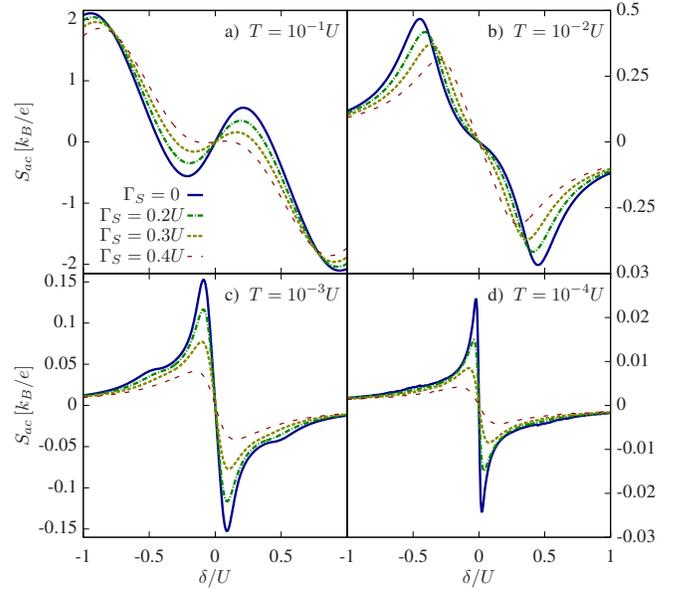}
\caption{\label{fig:Sacc-eps}
(color online) The thermopower
in the parallel configuration as a function
of $\delta$ for different $\Gamma_S$ and for
different temperatures in the case of finite spin accumulation.
The parameters are the same as in \fig{fig:AAP}.}
\end{figure}

\begin{figure}[t]
\centering
\includegraphics[width=1\columnwidth]{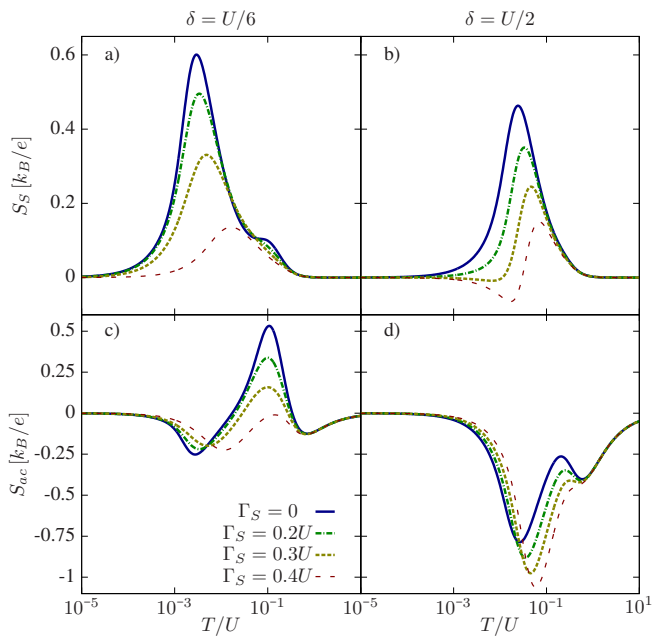}
\caption{\label{fig:Sacc-T}
(color online) The spin thermopower (a,b) and thermopower (c,d)
in the parallel configuration as a function of temperature for different 
$\Gamma_S$ in the case of finite spin accumulation.
The left column corresponds to $\delta = U/6$,
while the right column to $\delta = U/2$.
The parameters are as in \fig{fig:AAP}.}
\end{figure}

For completeness, in \fig{fig:Sacc-eps} we show the detuning dependence
of the Seebeck coefficient for different temperatures and couplings $\Gamma_S$
in the case of finite spin accumulation in the leads.
Except for opposite sign, the dependence of
$S_{ac}$ on $\delta$, $T$ and $\Gamma_S$
is quite similar to the dependence of the spin Seebeck coefficient,
cf. Figs.~\ref{fig:SS-eps} and \ref{fig:Sacc-eps},
although some differences still appear.
First of all, for temperatures considered in \fig{fig:Sacc-eps},
$S_{ac}$ exhibits a nonmonotonic dependence on $\delta$,
contrary to $S_S$, which for $T/U=0.1$ changed rather monotonically.
Moreover, the dependence on $\Gamma_S$ for $S_{ac}$ is now
weaker as compared to $S_S$. Increasing the coupling to the superconductor
results only in quantitative changes, leading generally 
to the suppression of the thermopower $S_{ac}$, see \fig{fig:Sacc-eps}.

Finally, it is also interesting to study the temperature dependence
of $S_S$ and $S_{ac}$ in the parallel configuration for different coupling $\Gamma_S$,
see \fig{fig:Sacc-T}. The left (right) column corresponds to
$\delta = U/6$ ($\delta = U/2$). Let us first consider the case of $\delta = U/6$,
when the system is in the Coulomb blockade regime.
For very high ($T > U$) or very low ($T<T_K$) temperatures,
both $S_S$ and $S_{ac}$ tend to zero for all values of $\Gamma_S$, see \figs{fig:Sacc-T}(a) and (c).
However, for intermediate temperatures, $T\sim T_K$, the spin Seebeck coefficient exhibits a maximum, 
the height of which diminishes with increasing $\Gamma_S$. The temperature
at which the maximum occurs increases when the coupling to the superconductor
is stronger. In addition, for $\Gamma_S=0$ there is also a 
small local maximum for $T\approx \Gamma$, which however 
merges with the large peak when $\Gamma_S$ is increased, see \fig{fig:Sacc-T}(a).
Contrary to $S_S$, the temperature dependence of $S_{ac}$ reveals
two sign changes. With increasing $T$, $S_{ac}$ first drops to a local minimum
for $T\sim T_K$, then changes sign and reaches a local maximum for $T\sim\Gamma$
to drop again for $T\sim U$ with another sign change, see \fig{fig:Sacc-T}(c).
When increasing the coupling to the superconductor, the temperature at which
the first minimum occurs increases, while the positions of other extrema are rather unchanged.
Moreover, the overall magnitude of $S_{ac}$ becomes generally suppressed with increasing $\Gamma_S$,
which is especially visible for $T\sim \Gamma$, see \fig{fig:Sacc-T}(c).

For larger detuning, $\delta=U/2$, the system is on resonance for $\Gamma_S=0$.
The temperature dependence of the spin thermopower displays then
a single maximum for temperatures of the order of the coupling $\Gamma$.
With increasing the coupling to the superconductor, this maximum
transforms into a local minimum and $S_S$ exhibits a sign
change when increasing temperature, see \fig{fig:Sacc-T}(b). This is opposite to $S_{ac}$,
which is now negative in the whole range of temperatures, irrespective of $\Gamma_S$,
see \fig{fig:Sacc-T}(d). The Seebeck coefficient has two local minima for $T\sim\Gamma$ and $T\sim U$,
separated by a local maximum, which merge together with increasing $\Gamma_S$
into a large single minimum for $T\sim\Gamma$.
Consequently, the proximity of the superconductor generally enhances the
Seebeck coefficient $S_{ac}$. This behavior is opposite to that of $S_S$
and $S_{ac}$ in the case of $\delta  = U/6$ discussed above, for which the proximity effect
led to a general suppression of the (spin) thermopower.

Finally, we would like to note that although the range of temperatures studied 
in \fig{fig:Sacc-T} may be slightly too large to assure
that the description based on effective Hamiltonian
in the limit of large superconducting gap is reasonable,
we showed the data at high temperatures $T\gtrsim U$ for completeness and consistency.
Nevertheless, the most interesting behavior of the Seebeck and spin Seebeck coefficients
discussed above occurs in the temperature range where the assumptions used are correct.

\section{Concluding remarks}

In the present paper we analyzed the electric and thermoelectric transport properties
of Kondo-correlated quantum dots coupled to the left and right ferromagnetic leads and
additionally coupled to one superconducting lead. In such hybrid devices, transport 
characteristics are determined by the interplay of ferromagnetic-contact induced exchange field,
the superconducting proximity effect and correlations leading to the Kondo effect.
By using the full density-matrix numerical renormalization group method, we
determined the dot's spectral function, linear electric and thermal conductances,
the TMR and the (spin) Seebeck coefficient for different temperatures, level positions
and couplings to the superconductor in the limit of large superconducting gap.
We showed that the superconducting proximity effect may considerably 
affect the exchange field, which is a function of Andreev bound state energies.
For the exchange field, we provided an approximative analytical formula
that agrees well with the NRG calculations.
The exchange field leads to a spin-splitting of the dot level,
which can suppress the Kondo resonance. We demonstrated that
increasing the coupling to the superconductor may raise the Kondo temperature
and partially restore the exchange-field-split Kondo resonance.
This subtle competition between ferromagnetic and superconducting
proximity effects is clearly visible in the corresponding temperature and level detuning
dependence of both the electric and thermoelectric transport coefficients of the system.

\acknowledgements
The discussions with J. Barna\'s, K. Bocian, W. Rudzi\'nski and P. Trocha are gratefully acknowledged.
We also thank J. Barna\'s and P. Trocha for critical reading of this manuscript.
This work is supported by the `Iuventus Plus' project No. IP2011 059471 for years 2012-2014
and the National Science Center in Poland as the Project No. DEC-2012/04/A/ST3/00372.
This research was also supported by a Marie Curie FP-7-Reintegration-Grants
(grant No. CIG-303 689) within the $7^{\rm th}$ European Community Framework Programme.


\begin{thebibliography}{9}

\bibitem{nato}
{\it Single Charge Tunneling: Coulomb Blockade Phenomena in
Nanostructures}, NATO ASI Series B: Physics 294, ed. by
H.~Grabert, M.H.~Devoret (Plenum Press, New York 1992).

\bibitem{schon}
{\it Mesoscopic Electron Transport}, ed. by L.L.~Sohn,
L.P.~Kouwenhoven, G.~Sch\"on (Kluwer, Dordrecht 1997).

\bibitem{loss02}
{\it Semiconductor Spintronics and Quantum Computation}, ed. by
D.D.~Awschalom, D.~Loss, and N.~Samarth (Springer, Berlin 2002).

\bibitem{andregassen10}
S Andergassen, V. Meden, H. Schoeller, J Splettstoesser and M R Wegewijs,
Nanotechnology {\bf 21}, 272001 (2010).


\bibitem{kondo}
J. Kondo, Prog. Theor. Phys. {\bf 32}, 37 (1964).

\bibitem{hewson_book}
A. C. Hewson, {\it The Kondo Problem to Heavy Fermions} (Cambridge
University Press, Cambridge, 1993).

\bibitem{goldhaber-gordon_98}
D. Goldhaber-Gordon, H. Shtrikman, D. Mahalu, D. Abusch-Magder, U.
Meirav, and M. A. Kastner, Nature (London) {\bf 391}, 156 (1998).

\bibitem{cronenwett_98}
S. Cronenwett, T. H. Oosterkamp, and L. P. Kouwenhoven, Science
{\bf 281}, 182 (1998).

\bibitem{buitelaarPRL03}
M. R. Buitelaar, W. Belzig, T. Nussbaumer, B. Babic, C. Bruder, and C. Sch\"onenberger,
Phys. Rev. Lett. {\bf 91}, 057005 (2003).

\bibitem{jorgensenPRL06}
H. I. Jorgensen, K. Grove-Rasmussen, T. Novotny,  K. Flensberg, and P. E. Lindelof,
Phys. Rev. Lett. {\bf 96}, 207003 (2006).

\bibitem{damNature06}
J. A. van Dam, Y. V. Nazarov, E. P. A. M. Bakkers, S. De Franceschi, L. P. Kouwenhoven,
Nature {\bf 442}, 667 (2006).

\bibitem{hakonenPRL07}
T. Tsuneta, L. Lechner, and P. J. Hakonen, Phys. Rev. Lett. {\bf 98}, 087002 (2007).

\bibitem{buitelaarPRL02}
M. R. Buitelaar, T. Nussbaumer, and C. Sch\"onenberger, Phys. Rev. Lett. {\bf 89}, 256801 (2002).

\bibitem{hechtJPCM08}
T. Hecht, A. Weichselbaum, J. von Delft, R. Bulla,
J. Phys.: Condens. Matter {\bf 20}, 275213 (2008).


\bibitem{pasupathy04}
A. N. Pasupathy, R. C. Bialczak, J. Martinek, J. E. Grose, L. A. K. Donev,
P. L. McEuen, and D. C. Ralph, Science {\bf 306}, 86 (2004).

\bibitem{barnasJPCM08}
J. Barna\'s and I. Weymann, J. Phys.: Condens. Matter {\bf 20},
423202 (2008).

\bibitem{martinekPRL03}
J. Martinek, M. Sindel, L. Borda, J Barna\'s, J. K\"onig, G. Sch\"on, J. von Delft,
Phys. Rev. Lett. {\bf 91}, 247202 (2003).

\bibitem{hauptmannNatPhys08}
J. Hauptmann, J. Paaske, P. Lindelof, Nature Phys. {\bf 4}, 373
(2008).

\bibitem{gaassPRL11}
M. Gaass, A. K. H\"uttel, K. Kang, I. Weymann, J. von Delft, and Ch. Strunk,
Phys. Rev. Lett. {\bf 107}, 176808 (2011).

\bibitem{sindelPRB07}
M. Sindel, L. Borda, J. Martinek, R. Bulla, J. K\"onig, G.
Sch\"on, S. Maekawa, and J. von Delft,
Phys. Rev. B {\bf 76}, 045321 (2007).

\bibitem{weymannPRB11} 
I. Weymann, Phys. Rev. B {\bf 83}, 113306 (2011).

\bibitem{wojcikJPCM13}
K. P. W\'ojcik, I. Weymann, and J. Barna\'s,
J. Phys.: Cond. Matter {\bf 25}, 075301 (2013).

\bibitem{hofstetterPRL10}
L. Hofstetter, A. Geresdi, M. Aagesen, J. Nygard, C. Sch\"onenberger, S. Csonka,
Phys. Rev. Lett. {\bf 104}, 246804 (2010).

\bibitem{zhuPRB01}
Y. Zhu, Q.-F. Sun, T.-H. Lin, Phys. Rev. B {\bf 65}, 024516 (2001).
\bibitem{fengPRB03}
J.-F. Feng, S.-J. Xiong, Phys. Rev. B {\bf 67}, 045316 (2003).
\bibitem{caoPRB04}
X. F. Cao, Y. Shi, X. Song, S. Zhou, H. Chen, Phys. Rev. B {\bf 70}, 235341 (2004).
\bibitem{pengZhang}
P. Zhang and Y. -X. Li, J. Phys.: Condens. Matter {\bf 21}, 095602 (2009).
\bibitem{konig09}
D. Futterer, M. Governale, M. G. Pala, and J.
K\"onig, Phys. Rev. B {\bf 79}, 054505 (2009).
\bibitem{konigPRB10}
B. Sothmann, D. Futterer, M. Governale,
and J. K\"onig, Phys. Rev. B {\bf 82}, 094514 (2010).
\bibitem{siqueiraPRB10}
E. C. Siqueira, G. G. Cabrera, Phys. Rev. B {\bf 81}, 094526 (2010).
\bibitem{wysokinskiJPCM}
K. I. Wysoki\'nski J. Phys.: Condens. Matter {\bf 24}, 335303 (2012).
\bibitem{bocian13}
K. Bocian, W. Rudzi\'nski, Eur. Phys. J. B {\bf 86}, 439 (2013).
\bibitem{weymann14}
I. Weymann and P. Trocha, Phys. Rev. B {\bf 89}, 115305 (2014).

\bibitem{WilsonRMP75}
K. G. Wilson, Rev. Mod. Phys. {\bf 47}, 773 (1975).

\bibitem{BullaRMP08}
R. Bulla, T. A. Costi, and T. Pruschke, Rev. Mod. Phys. {\bf 80},
395 (2008).

\bibitem{WeichselbaumPRL07}
A. Weichselbaum and J. von Delft, Phys. Rev. Lett. {\bf 99},
076402 (2007).

\bibitem{FlexibleDMNRG}
We used an open-access Budapest NRG code,
http://www.phy.bme.hu/dmnrg/;
O. Legeza, C. P. Moca, A. I. T\'{o}th, I. Weymann, G. Zar\'{a}nd,
arXiv:0809.3143 (2008) (unpublished).


\bibitem{trochaPRB12}
P. Trocha, J. Barna\'s,  Phys. Rev. B {\bf 85}, 085408 (2012).

\bibitem{hick}
L. D. Hicks and M. S. Dresselhaus, Phys. Rev. B {\bf 47}, 16631 (1993).

\bibitem{costiPRB10}
T. A. Costi and V. Zlatic, Phys. Rev. B {\bf 81}, 235127 (2010).

\bibitem{rejecPRB12}
T. Rejec, R. Zitko, J. Mravlje, and A. Ramsak, Phys. Rev. B {\bf 85}, 085117 (2012).

\bibitem{weymannPRB13}
I. Weymann and J. Barna\'s, Phys. Rev. B {\bf 88}, 085313 (2013).

\bibitem{chirlaPRB14}
R. Chirla and C. P. Moca, Phys. Rev. B {\bf 89}, 045132 (2014).

\bibitem{nagamatsu01} 
J. Nagamatsu, N. Nakagawa, T. Muranaka, Y. Zenitani, and J. Akimitsu, 
Nature {\bf 410}, 63 (2001).

\bibitem{heinrich13} 
B. W. Heinrich, L. Braun, J. I. Pascual and K. J. Franke,
Nature Phys. {\bf 9}, 765 (2013).

\bibitem{rozhkov} A. V. Rozhkov and D. P. Arovas, Phys. Rev. B {\bf 62}, 6687 (2000).

\bibitem{karraschPRB09}
C. Karrasch and V. Meden, Phys. Rev. B {\bf 79}, 045110 (2009).

\bibitem{mengPRB09}
T. Meng, S. Florens, and P. Simon, Phys. Rev. B {\bf 79}, 224521 (2009).

\bibitem{julliere}
M. Julliere, Phys. Lett. A {\bf 54}, 225 (1975).

\bibitem{zitko}
R. Zitko and T. Pruschke, Phys. Rev. B {\bf 79}, 085106 (2009).

\bibitem{oliveira}
V. L. Campo and L. N. Oliveira, Phys. Rev. B {\bf 72}, 104432 (2005).

\bibitem{friedel}
J. Friedel, Can. J. Phys. {\bf 34}, 1190 (1956).

\bibitem{weymannPRB05}
I. Weymann, J. K\"onig, J. Martinek, J. Barna\'s, and G. Sch\"on,
Phys. Rev. B {\bf 72}, 115334 (2005).



\end{thebibliography}
\end{document}